\documentclass[draftcls]{IEEEtran}
\usepackage[utf8]{inputenc}
\usepackage{mathtools,amssymb}
\usepackage{graphicx}
\usepackage{mathrsfs}
\usepackage[dvipsnames]{xcolor}
\colorlet{bvb}{Red}
\colorlet{wz}{blue}
\colorlet{bvb2}{orange}

\DeclareMathAlphabet\mathbfcal{OMS}{cmsy}{b}{n}

\DeclareMathOperator*{\argmin}{arg\,min}

\DeclareRobustCommand*{\IEEEauthorrefmark}[1]{%
  \raisebox{0pt}[0pt][0pt]{\textsuperscript{\footnotesize #1}}%
}
\begin{document}

\title{Transfer Learning Capabilities of Untrained Neural Networks for MIMO CSI Recreation}

\author{
\IEEEauthorblockN{\textit{Brenda Vilas Boas\IEEEauthorrefmark{1}$^,$\IEEEauthorrefmark{2}, Wolfgang Zirwas\IEEEauthorrefmark{1}}, \textit{Martin Haardt\IEEEauthorrefmark{2}}}

\IEEEauthorblockA{\IEEEauthorrefmark{1}Nokia, Germany \\\
\IEEEauthorrefmark{2}Ilmenau University of Technology, Germany \\\
{brenda.vilas\_boas@nokia.com, wolfgang.zirwas@nokia-bell-labs.com, martin.haardt@tu-ilmenau.de} }
}
\maketitle

\begin{abstract}
    Machine learning (ML) applications for wireless communications have gained momentum on the standardization discussions for 5G advanced and beyond. 
    One of the biggest challenges for real world ML deployment is the need for labeled signals and big measurement campaigns. 
    To overcome those problems, we propose the use of untrained neural networks (UNNs) for MIMO channel recreation/estimation and low overhead reporting. The UNNs learn the propagation environment by fitting a few channel  measurements and we exploit their learned prior to provide higher channel estimation gains. 
    Moreover, we present a UNN for simultaneous channel recreation for multiple users, or multiple user equipment (UE) positions, in which we have a trade-off between the estimated channel gain and the number of parameters. 
    Our results show that transfer learning techniques are effective in accessing the learned prior on the environment structure as they provide higher channel gain for neighbouring users. 
    Moreover, we indicate how the under-parameterization of UNNs can further enable low-overhead channel state information (CSI) reporting.     
    
\end{abstract}


\section{Introduction}

Standardization discussions for the next generation of wireless communications have  started
and artificial 
intelligence and machine learning (AI/ML) solutions are 
being considered, especially for next generation radio access network (NG-RAN) and air interfaces, i.e., $3$GPP release $18$ workshop~\cite{213gppRel18} may lead to a study item on AI/ML applications for the PHY layer. 
Moreover, reducing the overall power consumption of the system is a target for $5$G advanced and $6$G. 
In this context, we assume the use of a digital twin environment~\cite{19NokiaDigital} where learning of the AI/ML solutions takes place, and 
can later aid planning, deployment and management of wireless networks. 
In order to leverage the potential of a digital twin for the environment, 
full knowledge of the channel state information (CSI) is needed such that most of the real propagation effects 
can be represented. 
Here, we propose a ML solution based on 
untrained neural networks (UNNs) 
for channel recreation/estimation at the initial 
operation phase (day zero) where not many CSI measurements are available. 
For instance, we can design a UNN to learn the environment characteristics based on a single time snapshot channel measurement. However, it is beneficial for the UNN to collect some few channel measurements over time.
Our method is an enabler for PHY functionalities and other AI/ML methods which need CSI as labels, e.g., fingerprinting, CSI-prediction, and multi-modal data-aided networks (lidar, radar, environment images, etc.)~\cite{21Joint6G}. 
Moreover, our proposed ML solution can further leverage full CSI reporting between the user equipment (UE) and the base station (BS).




Untrained neural networks (UNNs) were first proposed in~\cite{18LempitskyDIP} where the term untrained relates to the fact that a huge data collection for training is not needed. 
Instead, the neural network can be fitted to a single data sample. Therefore, the updates of gradient descent move from the training phase to the inference phase. 
The solution of inverse problems, such as denoising, is feasible because the structure of deep convolutional networks act as a prior to an image-like signal~\cite{18LempitskyDIP}. The work in~\cite{18HeckelDeep} goes further and proposes a underparameterized \textit{deep decoder} architecture for UNNs. 
For wireless systems, this means we can fit a UNN to directly estimate the multidimensional wireless channel based on a small measurement campaign, i.e., a few time snapshots, without the need of `true' labels.

The application of UNNs for real inverse problems is quite recent. For instance, the work in~\cite{21DarestaniMRI} proposes to use a convolutional \textit{deep decoder} UNN to accelerate magnetic resonance imaging which has superior performance than total variation norm minimization. In \cite{20BaleviUNN}, a MIMO channel estimator using UNN is proposed to overcome pilot contamination. However, 
the authors evaluate the performance
using the LTE-EPA channel model which does not reveal one of the main advantages of UNN channel estimators: prior knowledge on the propagation environment which is stored in the UNN structure.

The storage of prior knowledge, the short 
data-collection phase and the small 
number of parameters of UNNs have motivated us to further investigate their use for CSI recreation. 
We refer to channel recreation instead of channel estimation to emphasize that channel estimation gain is not the main goal of our UNN architecture design. 
The UNN is optimized to estimate the wireless channel with at least the same signal to noise ratio (SNR) as the corresponding channel measurement. 
Therefore, the UNN learns prior knowledge about the propagation environment. 
In this work, we propose to take advantage of the learned prior knowledge by means of transfer learning~\cite{20BozinovskiTransfer}. 
In addition, we propose to expand the UNN structure to be able to recreate simultaneously the channel of multiple neighboring UEs. 
Moreover, we indicate how the UNN structure can be exploited for low-overhead full CSI reporting. 
In contrast to prior art,
we evaluate the performance of our UNN channel estimators on geometrically modeled channels which better represent environment specific fading characteristics.

In this paper, Section~\ref{sec:system} presents the wireless propagation environment, Section~\ref{sec:2D-UNN} presents the UNN based single user CSI estimator and the transfer learning approach. After that, Section~\ref{sec:3D-UNN} presents the UNN based simultaneous CSI estimator for multiple UEs. Finally, Section~\ref{sec:symul} present our simulation results and Section~\ref{sec:conclusion} concludes our paper.

Regarding the notation, $a$, $\mathbf{a}$, $\mathbf{A}$ and $\mathbfcal{A}$ represents, respectively, scalars, column vectors, matrices and $D$-dimensional tensors. The superscript $^T$, denotes transposition. 
For a tensor $\mathbfcal{A} \in \mathbb{C}^{M_1 \times M_2 \times \dots M_D}$, $M_d$ refers to the tensor dimension on the $d^\mathrm{th}$ mode.
A $d$-mode unfolding of a tensor is written as $[\mathbfcal{A}]_{(d)} \in \mathbb{C}^{M_d \times M_{d+1} \dots M_D M_1 \dots M_{d-1}}$ where all $d$-mode vectors are aligned as columns of a matrix. The $d$-mode vectors of $\mathbfcal{A}$ are obtained by 
varying the $d^\mathrm{th}$ index from $1$ to $M_d$ and keeping all other indices fixed.
Moreover, $\mathbfcal{A} \times_d \mathbf{U}$ is the $d$-mode product between a $D$-way tensor $\mathbfcal{A} \in \mathbb{C}^{M_1 \times M_{2} \dots \times M_D}$ 
and a matrix $\mathbf{U} \in \mathbb{C}^{J \times M_d}$. The $d$-mode product is computed by multiplying $\mathbf{U}$ with
all $d$-mode vectors of
$\mathbfcal{A}$.

\section{System and Channel Models}
\label{sec:system}

\begin{figure}[tb!]
    \centering
    \includegraphics[width=\columnwidth]{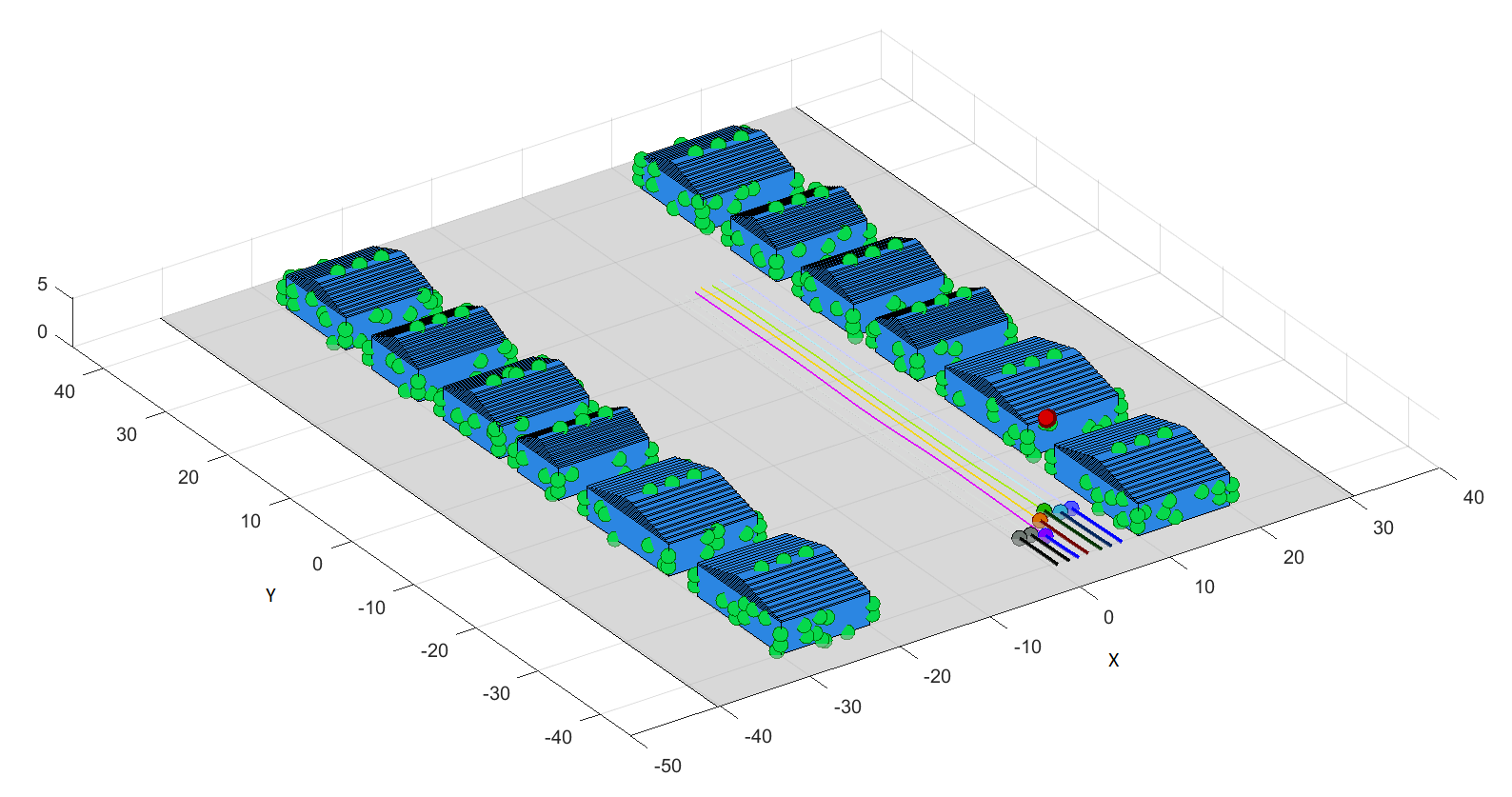}
    \caption{Propagation environment simulated in IlmProp. Numbered axis represents the positions in meters.}
    \label{fig:IlmProp}
\end{figure}

For the problem of channel recreation and transfer learning with UNNs, 
we consider an urban environment with a fixed base station (BS) equipped with an uniform rectangular array (URA) containing $N_\mathrm{ant}$ antenna elements, moving user equipments (UEs) with single antennas, operating with $N_\mathrm{sub}$ OFDM subcarriers, and collecting $N_\mathrm{sp}$ time snapshots. 
This scenario was modeled with the IlmProp, a geometry based channel simulator developed at Ilmenau University of Technology~\cite{05GaldoGeometry}.  
Figure~\ref{fig:IlmProp} presents the urban environment with the BS represented by a red sphere, buildings in blue cuboids, scatters in green spheres, and seven UEs moving on a linear trajectory towards the BS. In addition, the numbered $x-y$-plane represent the position on the map in meters.

The channels generated by IlmProp are the ground truth values $\mathbfcal{H}_\mathrm{sim} ~\in~\mathbb{C}^{N_\mathrm{sub} \times N_\mathrm{sp} \times N_\mathrm{ant}}$. 
From those, we derive the noisy channel measurements $\mathbfcal{H}_\mathrm{mes}^C~\in~\mathbb{C}^{N_\mathrm{sub} \times N_\mathrm{sp} \times N_\mathrm{ant}}$ as  
\begin{equation}
    \mathbfcal{H}_\mathrm{mes}^C = \mathbfcal{H}_\mathrm{sim} + \mathbfcal{N}, 
    \label{eq:mes}
\end{equation}
where $\mathbfcal{N} \in \mathbb{C}^{N_\mathrm{sub} \times N_\mathrm{sp} \times N_\mathrm{ant}}$ is a zero mean circularly symmetric complex Gaussian noise process.
The $\mathbfcal{H}_\mathrm{mes}^C$ are further used to recreate the CSIs. 

After we derive the optimum weights for a UNN to recreate the CSIs of $N_\mathrm{sp}$ different locations, we can send the UNN weights together with its structural details (such as random input rule, and number of upsampling operations) from UE to BS. Hence, the BS would be able to recreate the same CSI as the reporting UE. Since UNNs are under-parameterized, we achieve compression of the full-CSI ($\mathbfcal{H}_\mathrm{mes}^C$). In this work, we present how to achieve CSI compression by different UNN structures that take advantage of the channel correlation between neighboring UEs. Nonetheless, as we discuss in Section~\ref{subsec:transfer}, it is possible to apply other compression schemes on top of the UNN weights to achieve an even higher compression rate.
The use of UNN structures for CSI reporting is an alternative to variational auto-encoder (VAE) solutions, which are trained to generate codewords that represent the channels. The motivation for our method is to inherently also learn the environment, i.e., to generate a ML based digital twin or mirror world of the environment.

\section{Single user CSI estimation and Transfer Learning}
\label{sec:2D-UNN}

Due to the claim in~\cite{18LempitskyDIP} that the network structure stores the prior-information, we aim to access this prior by means of transfer learning. 
In wireless channel estimation, access to prior information can provide a channel estimation gain if the correlation between the channels is considered. 
Physically, neighboring UEs are favorable candidates as they experience similar propagation effects in an environment. 
Hence, a UNN is learning the propagation environment while fitting the channel measurements, without any direct knowledge of the environment map. 
In this section, we present the UNN channel estimator based on the \textit{deep decoder} architecture~\cite{18HeckelDeep} for a single UE. 
We introduce the data pre-processing, the UNN architecture, and how gradient descent is used to update the weights.       
Moreover, we propose to use transfer learning to take advantage of the stored prior in the UNN weights.


\begin{figure}[tb!]
    \centering
    \includegraphics[width=\columnwidth]{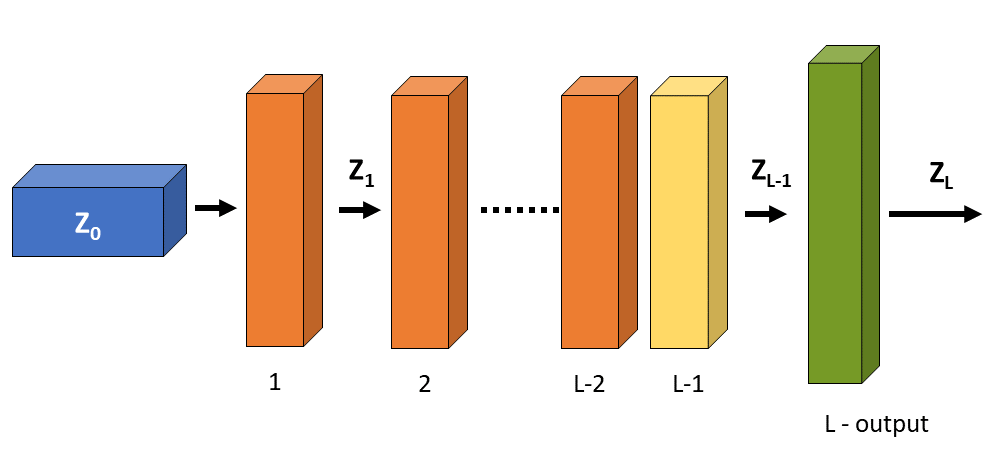}
    \caption{General structure of a untrained neural network (UNN) $P$ that maps $\mathbfcal{Z}_{L} = P(\mathbfcal{K}, \mathbfcal{Z}_0)$. After $I$ gradient iterations, the  wireless channel is estimated as $\mathbfcal{H}_\mathrm{est} = P(\mathbfcal{K}^*, \mathbfcal{Z}_0)$.}
    \label{fig:unn}
\end{figure}

\subsection{Data pre-processing for UNN}

The input signal to a UNN is a random noise seed $\mathbfcal{Z}_0 \in \mathbb{R}^{b \times c \times k_0}$, where $b = N_\mathrm{sub}/2^{L-2}$, $c = N_\mathrm{sp}/2^{L-2}$,  $k_0$ is the depth of the random seed which is a hyper-parameter, and $L$ is the number of layers. The input tensor 
$\mathbfcal{Z}_0$ is drawn from a uniform distribution $U(-a,+a)$ defined on the interval $[-a, +a]$ and kept fixed during the iterations to update the gradient descent.   
The measured channel $\mathbfcal{H}_\mathrm{mes}^C$ is preprocessed as 
\begin{itemize}
    \item Each time snapshot within $\mathbfcal{H}_\mathrm{mes}^C$ is normalized by its Frobenius norm, and then multiplied by a scaling factor to ease convergence.
    \item $\mathbfcal{H}_\mathrm{mes}^C \in \mathbb{C}^{N_\mathrm{sub} \times N_\mathrm{sp} \times N_\mathrm{ant}}$ is rearranged by concatenating $\mathfrak{Re} \{ \mathbfcal{H}_\mathrm{mes}^C \}$ and $\mathfrak{Im} \{ \mathbfcal{H}_\mathrm{mes}^C \}$ in the dimension corresponding to the antenna elements. 
\end{itemize}
After those operations, $\mathbfcal{H}_\mathrm{mes} \in \mathbb{R}^{N_\mathrm{sub} \times N_\mathrm{sp} \times 2N_\mathrm{ant}}$ is directly used to compute the cost function. 

\subsection{UNN architecture for single UE CSI estimation}
\label{subsec:singleCSIUNN}
The UNN \textit{deep decoder} architecture is a composition of $L$ layers which are of three types: $(L-2)$ inner layers, one pre-output layer $(L-1)$ and one output layer $(L)$.
Figure~\ref{fig:unn} shows a generic organization of those layers, the random noise seed $\mathbfcal{Z}_0$ in blue, the inner layers in orange, the pre-output layer in yellow, and the output layer in olive. 
All the layer types contain convolutional filters $\mathbfcal{W}_l \in \mathbb{R}^{1 \times 1 \times k_{l-1} \times k_{l}}$ where $l = \{1, 2, \ldots L \}$, $k_{l-1}$ and $k_{l}$ are hyper-parameters which define the number of filters on the respective
$(l-1)^\mathrm{th}$ and $l^\mathrm{th}$ layers. However, the types of layers differ
with respect to the upsampling computation and the operation of batch normalization. 
The inner layers contain linear and non-linear operations. The linear computation consists of a convolutional filter $\mathbfcal{W}_l$ and a bilinear upsampling operation, where $\mathbf{A}_l \in \mathbb{R}^{{2^l}b \times 2^{l-1}b}$ and  $\mathbf{C}_l \in \mathbb{R}^{2^{l}c \times 2^{l-1}c}$ are the one dimensional linear upsampling matrices in the subcarrier and time snapshots dimensions, respectively. 
Next, the rectifier linear unit (ReLu) activation function is applied, and a batch normalization is computed
per $k_l$ filter as
\begin{equation}
    \mathrm{BatchNorm}(\mathbfcal{Z}_{lj}) = \frac{\mathbfcal{Z}_{lj} - \mathrm{mean}(\mathbfcal{Z}_{lj})}{\sqrt{\mathrm{var}(\mathbfcal{Z}_{lj})}} \gamma_{lj} + \beta_{lj}, 
\end{equation}
where $j=[1, 2, \ldots, k_l]$, mean and variance (var) are computed among the batch samples~\cite{15IoffeBatch} which for UNN is one. The trainable parameters of the $\mathrm{BatchNorm}$ operation are 
$\mathbf{R}_l = [ \boldsymbol{ \gamma}_l, \boldsymbol{ \beta}_l ] \in \mathbb{R}^{k_l \times 2}$. 
For instance, the output of the first inner layer $\mathbfcal{Z}_1 $ can be written as
\begin{equation}
    \mathbfcal{Z}_1 = \mathrm{BatchNorm} (\mathrm{ReLu}(\mathbfcal{Z}_0 \times_3 [\mathbfcal{W}_1]_{(4)} \times_1 \mathbf{A}_1 \times_2 \mathbf{C}_1^T)),
\end{equation}
where 
$[\mathbfcal{W}_1]_{(4)}$ is the $4$-mode unfolding of the convolutional filters operating at the antenna elements dimension.  
The pre-output layer differs from the inner layers because it does not apply upsampling. Hence, it can be written as
\begin{equation}
    \mathbfcal{Z}_{L-1} = \mathrm{BatchNorm} (\mathrm{ReLu}(\mathbfcal{Z}_{L-2} \times_3 [\mathbfcal{W}_{L-1}]_{(4)})).
\end{equation}
After that, the output layer is used to 
adjust the range of values as well as the size expected in the output $k_L = 2N_\mathrm{ant}$, such that
\begin{equation}
    \mathbfcal{Z}_{L} = \mathrm{TanH}(\mathbfcal{Z}_{L-1} \times_3 [\mathbfcal{W}_{L}]_{(4)}),
\end{equation}
where $\mathbfcal{W}_{L} \in \mathbb{R}^{1 \times 1 \times k_{l-1} \times 2N_\mathrm{ant}}$, and $\mathrm{TanH}$ is the hyperbolic tangent activation function.
Since the upsamplig operations are pre-defined, the trainable parameters reduce to the convolutional filters $\mathbfcal{W}_{l}$ and the regularization parameters $\mathbf{R}_l$ of the batch normalization operation. Therefore, $\mathbfcal{K}_l = \{\mathbfcal{W}_{l}, \mathbf{R}_l\}$ is the set of trainable parameters 
of the $l^\mathrm{th}$ layer, and $\mathbfcal{K}$ refers to all the trainable parameters
of the $L$ layers.

\subsection{Updating the weights of a UNN}
The UNN is a model $P: \mathbb{R}^N \rightarrow \mathbb{R}^{N_\mathrm{sub}N_\mathrm{sp}2N_\mathrm{ant}}$ where $N<N_\mathrm{sub}N_\mathrm{sp}2N_\mathrm{ant}$ is its total number of parameters used to map the input tensor $\mathbfcal{Z}_0$ to the output tensor $\mathbfcal{Z}_L$ as $\mathbfcal{Z}_L  = P(\mathbfcal{K},\mathbfcal{Z}_0)$. 

The loss function of the UNN is the mean squared error (MSE) which is computed as 

\begin{equation}
    \mathcal{L}(\mathbfcal{K}) = \mathbb{E} \{\parallel P(\mathbfcal{K}, \mathbfcal{Z}_0) - \mathbfcal{H}_\mathrm{mes} \parallel^2_F \}.
\end{equation}
The trainable parameters $\mathbfcal{K}$ are updated by $I$ gradient descent iterations such that 
\begin{equation}
    \mathbfcal{K}^* = \underset{\mathbfcal{K}}{\argmin}~ \mathcal{L}(\mathbfcal{K}), ~ \mathrm{and} ~
    \mathbfcal{H}_\mathrm{est} = P(\mathbfcal{K}^*, \mathbfcal{Z}_0).
\end{equation}
Therefore, $\mathbfcal{H}_\mathrm{est}$ is the channel estimated for a single UE by the UNN $P$ when optimized for a specific $\mathbfcal{H}_\mathrm{mes}$. 
Since there is no big data collection phase, UNNs do not have generalization capabilities.  

\subsection{Transfer Learning for UNNs}
\label{subsec:transfer}
Since the input tensor $\mathbfcal{Z}_0$ is fixed, the mapping weights $\mathbfcal{K}^*$ are only able to recover the
considered $\mathbfcal{H}_\mathrm{mes}$  during the $I$ gradient iterations. In addition, a change in the construction of the random seed makes the estimation task impossible. For instance, if we generate $\mathbfcal{Z}_0'$ from a different seed number compared to $\mathbfcal{Z}_0$, $\mathbfcal{H}_\mathrm{est} \neq P(\mathbfcal{K}^*, \mathbfcal{Z}_0')$ the output of the UNN $P$ is something different from $\mathbfcal{H}_\mathrm{est}$.   
Nonetheless, according to~\cite{18LempitskyDIP},
the priors are stored in $\mathbfcal{K}^*$. Therefore, we propose to apply transfer learning in order to take advantage of the prior-knowledge for wireless channel estimation. 

Let us consider two neighboring UEs with the same number of antennas in a certain propagation environment, $\mathbfcal{H}_{1,\mathrm{mes}} \in \mathbb{R}^{N_\mathrm{sub} \times N_\mathrm{sp} \times 2N_\mathrm{ant}}$ and $\mathbfcal{H}_{2,\mathrm{mes}} \in \mathbb{R}^{N_\mathrm{sub} \times N_\mathrm{sp} \times 2N_\mathrm{ant}}$ are the measured wireless channel for each of them. In order to estimate the channel for $\mathbfcal{H}_{1,\mathrm{est}}$ UE $1$, the weights of the UNN estimator are initialized from random values $\mathbfcal{K}_{1,0}$ and iterated such that 
\begin{equation}
    \mathbfcal{K}_{1,0}^* = \underset{\mathbfcal{K}_{1,0}}{\argmin}~ \parallel P(\mathbfcal{K}_{1,0}, \mathbfcal{Z}_0) - \mathbfcal{H}_{1,\mathrm{mes}} \parallel^2_F, 
\end{equation}
and $\mathbfcal{H}_{1,\mathrm{est}} = P(\mathbfcal{K}_{1,0}^*, \mathbfcal{Z}_0)$. Here, we assume $\mathbfcal{K}_{1,0}^*$ is the set of projection tensors which operates sequentially over $\mathbfcal{Z}_0$ to reconstruct $\mathbfcal{H}_{1,\mathrm{est}}$. Next, we propose to estimate the channel of UE $2$ $\mathbfcal{H}_{2,\mathrm{est}}$ as
\begin{equation}
    \mathbfcal{K}_{2,1}^* = \underset{\mathbfcal{K}_{1,0}^*}{\argmin}~ \parallel P(\mathbfcal{K}_{1,0}^*, \mathbfcal{Z}_0) - \mathbfcal{H}_{2,\mathrm{mes}} \parallel^2_F, 
\end{equation}
where the weights are not initialized from random values $\mathbfcal{K}_{2,0}$, but from the weights of its neighbor, UE $1$ in this case. Hence, $\mathbfcal{H}_{2,\mathrm{est}} = P(\mathbfcal{K}_{2,1}^*, \mathbfcal{Z}_0)$ and the number of gradient iterations are the same for UE $1$ and UE $2$. This implies that we are constraining the gradient descent to search for a sub-space of solutions for UE $2$ close to the sub-space of UE $1$ since
\begin{equation}
\parallel \mathbfcal{K}_{1,0}^* - \mathbfcal{K}_{2,1}^* \parallel_F~ < ~\parallel \mathbfcal{K}_{1,0}^* - \mathbfcal{K}_{2,0}^* \parallel_F.     
\label{eq:filternorm}
\end{equation}
Moreover, if $\mathbfcal{H}_{1,\mathrm{mes}}$ and $\mathbfcal{H}_{2,\mathrm{mes}}$ are correlated, the channel gain obtained for
$\mathbfcal{H}_{2,\mathrm{est}}(\mathbfcal{K}_{2,1}^*) = P(\mathbfcal{K}_{2,1}^*, \mathbfcal{Z}_0)$ is higher than the gain of $\mathbfcal{H}_{2,\mathrm{est}}(\mathbfcal{K}_{2,0}^*) = P(\mathbfcal{K}_{2,0}^*, \mathbfcal{Z}_0)$ as $\mathbfcal{K}_{1,0}^*$ is a prior to $\mathbfcal{K}_{2,1}^*$.

This proposal is aligned with transfer learning~\cite{20BozinovskiTransfer} since we derive  knowledge for a 
$1^\mathrm{st}$
task (estimate $\mathbfcal{H}_{1,\mathrm{est}}$) and use it to solve a 
$2^\mathrm{nd}$
task (estimate $\mathbfcal{H}_{2,\mathrm{est}}$). Since neighboring wireless channels are likely to be correlated due to their propagation environment, the transfer learning is very suitable. Moreover, even if a channel 
estimation 
gain is not achieved, the distance between the weights are reduced which can be further leveraged by compression schemes when reporting the UNN-estimator parameters.   

\section{Simultaneous CSI estimation for multiple users}
\label{sec:3D-UNN}

In this section, we propose to extend the UNN-estimator to simultaneously estimate the channels of $M$ multiple neighboring UEs. 
Since neighboring UEs tend to have correlated channels, a UNN with three dimensional convolutional kernels can be optimized to find the weights that best fit the channel measurements. 
Despite the expansion of dimensions, the number of trainable parameters would not explode if there is enough correlation between the UEs considered. 
This is different from transfer learning, as here we start from a random initialization and output channels for $M$ UEs simultaneously. 
In the following subsections, we present the construction of the signals at the input and the output, as well as the architecture and its weights optimization.

\subsection{Data preparation}

Similar to the UNN for single user CSI estimation, the input signal is a random noise seed $\mathbfcal{Z}_0^M \in \mathbb{R}^{b \times c \times d \times k_0}$ where $b = N_\mathrm{sp}/2^n$, $c = N_\mathrm{sub}/2^n$, $d = M/2^n$, $k_0$ is the number of hyper-parameters of the input, $L$ is the number of layers, and $n = \{1, 2, \dots L-2\}$ is the number of layers with upsampling operation in the given dimension.
The input tensor 
$\mathbfcal{Z}_0^M$ is drawn from a uniform distribution $U(-a,+a)$ defined on the interval $[-a, +a]$ and kept fixed during the iterations to update the gradient descent.   
The channels measured for each $m^\mathrm{th}$ UE
$\mathbfcal{H}_{m,\mathrm{mes}} \in \mathbb{R}^{N_\mathrm{sp} \times N_\mathrm{sub} \times 2N_\mathrm{ant}}$, where $m=\{1, 2, \ldots, M\}$, are concatenated as $\mathbfcal{H}_\mathrm{mes}^M \in \mathbb{R}^{N_\mathrm{sp} \times N_\mathrm{sub} \times M \times 2N_\mathrm{ant}}$.   

\subsection{UNN architecture for multiple UE CSI estimation}
For estimating the CSI of multiple users, we propose to use a three dimensional convolutional kernel and apply trilinear upsampling to expand
the UNN architecture defined for single user CSI estimation
as in Section~\ref{subsec:singleCSIUNN}.
Hence, all the 
convolutional filters of the $L$ layers are $\mathbfcal{W}_l^M \in \mathbb{R}^{1 \times 1 \times 1 \times k_{l-1} \times k_{l}}$ where $l = \{1, 2, \ldots L \}$.
Moreover, the upsampling operation of the inner layers is defined by three one dimensional linear upsampling matrices: $\mathbf{A}_l \in \mathbb{R}^{{2^l}b \times 2^{l-1}b}$,  $\mathbf{C}_l \in \mathbb{R}^{2^{l}c \times 2^{l-1}c}$, and $\mathbf{D}_l \in \mathbb{R}^{2^{l}d \times 2^{l-1}d}$. 
The output of the first inner layer $\mathbfcal{Z}_1^M$, for example, is
\begin{equation}
\begin{split}
    \mathbfcal{Z}_1^M = \mathrm{BatchNorm} (\mathrm{ReLu}(\\
    \mathbfcal{Z}_0^M \times_4 [\mathbfcal{W}_1^M]_{(5)} \times_1 \mathbf{A}_1 \times_2 \mathbf{C}_1^T \times_3 \mathbf{D}_1)).
\end{split}
\end{equation}
Here, we point out that depending on the design choice of $N_\mathrm{sp}$, $N_\mathrm{sub}$ and $M$, we can disable the upsampling matrices accordingly. For instance, we should not consider $\mathbf{A}_l$ if just one time snapshot is available on the channel measurement. 
Next, 
the pre-output layer is
\begin{equation}
    \mathbfcal{Z}_{L-1}^M = \mathrm{BatchNorm} (\mathrm{ReLu}(\mathbfcal{Z}_{L-2}^M \times_4 [\mathbfcal{W}_{L-1}^M]_{(5)})),
\end{equation}
and the output layer is computed as
\begin{equation}
    \mathbfcal{Z}_{L}^M = \mathrm{TanH}(\mathbfcal{Z}_{L-1}^M \times_4 [\mathbfcal{W}_{L}^M]_{(5)}).
\end{equation}

\subsection{Optimization of the weights}
Let us define the UNN mapping function as $Q~:~\mathbb{R}^N~\rightarrow~\mathbb{R}^{N_\mathrm{sub}N_\mathrm{sp}M2N_\mathrm{ant}}$ where $N< N_\mathrm{sub}N_\mathrm{sp}M2N_\mathrm{ant}$ is the number of trainable parameters. Hence, the output of 
the UNN structure with three dimensional convolutional filters is 
$\mathbfcal{Z}_L^M = Q(\mathbfcal{K}^M,\mathbfcal{Z}_0^M)$ where the weights $\mathbfcal{K}^M$ are initialized from random values. 

The MSE is used as loss function to compute the gradient descent updates.
After $I$ gradient iterations, the optimum parameters are
\begin{equation}
    \mathbfcal{K}^{M*} = \underset{\mathbfcal{K}^M}{\argmin}~ \parallel Q(\mathbfcal{K}^M, \mathbfcal{Z}_0^M) - \mathbfcal{H}_\mathrm{mes}^M \parallel^2_F, 
\end{equation}
and the channel estimated for the $M$ users is $\mathbfcal{H}_\mathrm{est}^M = Q(\mathbfcal{K}^{M*}, \mathbfcal{Z}_0^M).$

\section{Simulations and Results}
\label{sec:symul}

As presented in Section~\ref{sec:system}, we use IlmProp to simulate a street canyon scenario as in Figure~\ref{fig:IlmProp} where UE $1$ is the closest to the building, and UE $7$ is the most distant.
Table~\ref{tab:SymIlm} presents the parameters we use to setup the simulation. For the measured channels, we vary the SNR in the range of $0$~dB to $20$~dB. 
The UNN channel estimators are simulated using Python and PyTorch without any graphics processing unit (GPU).
The results are compared according to the normalized mean squared error $\mathrm{NMSE} = \mathbb{E} \left\{ \frac{\parallel \mathbfcal{H}_\mathrm{est} - \mathbfcal{H}_\mathrm{sim} \parallel_F^2}{\parallel \mathbfcal{H}_\mathrm{sim} \parallel_F^2} \right\}$. 

\begin{table}[tb!]
    \centering
    \caption{Simulation parameters for IlmProp.}
    \label{tab:SymIlm}
    \resizebox{0.6\linewidth}{!}{%
    \begin{tabular}{|c|c|}
    \hline
         Parameter & Value \\ \hline
         Carrier frequency & 2.6~GHz \\ \hline
         Bandwidth & 20~MHz \\ \hline
         $N_\mathrm{sub}$ & 64 \\ \hline
         Velocity & [0.08, 0.14] ~m/s \\ \hline
         $N_\mathrm{sp}$ & 64 \\ \hline
         Total of UEs & 7 \\ \hline
         $N_\mathrm{ant}$ & 36 \\ \hline
    \end{tabular}}
\end{table}

First, we define the UNN structure to estimate single UE CSI channels. We choose to have four inner layers, each with
number of filters $k_{1:L-2}=64$ and both upsampling matrices, $\mathbf{A}_l$ and $\mathbf{C}_l$, activated in all inner layers. 
In addition, we use one pre-output layer with $k_{L-1} = 64$ filters and one output layer with $k_L = 72$ convolutional filters. Therefore, there are $L=6$ layers and the random noise seed $\mathbfcal{Z}_0~\in~\mathbb{R}^{4 \times 4 \times 64}$ is drawn from a uniform distribution as $U(-0.15, +0.15)$ where $k_0=64$. 
Second, 
the trainable parameters $\mathbfcal{K}$ are initialized from random values and $I=25000$ gradient updates are performed to
find the best $\mathbfcal{K}^*$ for each UE, separately. 
The design choice of the hyper-parameter $k$ and iterations $I$ were made to fulfill our requirement of CSI recreation, where the estimated channel should have the same or a higher SNR than the measured channel. 
The presented architecture for single UE CSI recreation contains $25728$ trainable parameters which correspond to  $17.45\% $ of the coefficients in  $\mathbfcal{H}_\mathrm{mes}^C~\in~\mathbb{C}^{64 \times 64 \times 36}$.

\begin{figure}[tb!]
    \centering
    \includegraphics[width=\columnwidth]{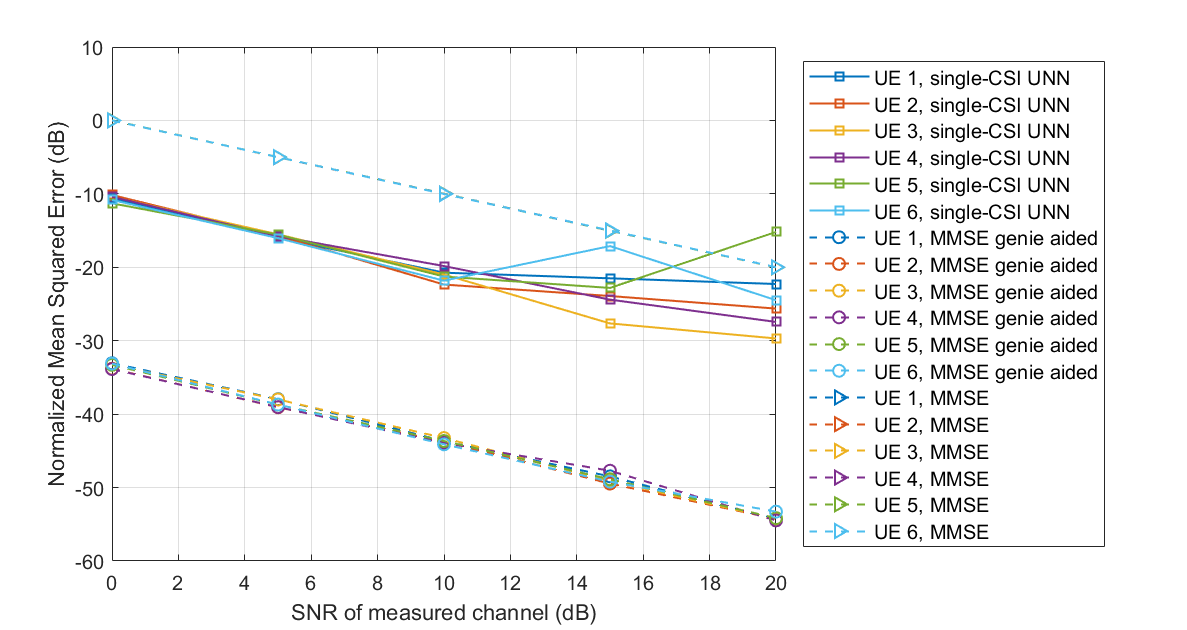}
    \caption{Simulation results for varying the measurement SNR and NMSE of the estimated channel. The same UNN architecture and number of gradient updates were used for all estimation tasks. For reference, we plot the performance of channel estimation with the MMSE estimator, without any noise reduction technique, and the genie aided MMSE estimator.}
    \label{fig:single}
\end{figure}

Figure~\ref{fig:single} presents the results for single UE CSI estimation where we apply the same 
UNN architecture ($L=6, ~k_{1:L-2}=64$), 
with random initialization, for the six UEs with varying SNRs.
For reference, we also plot the performance of the minimum mean squared error (MMSE) channel estimator without any noise reduction technique and the genie-aided MMSE estimator.
From Figure~\ref{fig:single}, we see that only UE$~5$ at SNR~$=~20~$dB is not able to meet the design requirement. 
This failure as well as the variance in channel estimation gain at SNR~$=20$~dB are mainly due to the different small fading characteristics of each UE. 
Moreover, for low SNRs ($0, 5, 10~\mathrm{dB}$),  there is a channel gain of about $10$~dB for all the six UEs. This is due to the fact that UNNs have a high impedance to fit noise~\cite{18LempitskyDIP}.         
Hence, the design of the UNN-estimator should be done at high-SNR values, where the noise levels are smaller, and it can be reused for low-SNR values.
This understanding is compatible with the results in~\cite{20BaleviUNN}. However, their authors prefer to reduce the number of hyper-parameters ($k$) and the number of iterations ($I$) when the measured SNR degrades. Such design choice further reduced the channel estimation gain.
Moreover,~\cite{20BaleviUNN} could not study the propagation environment effects on the UNN estimator performance since it applied channel models based on statistical distributions.

\begin{figure}[tp!]
    \centering
    \includegraphics[width=\columnwidth]{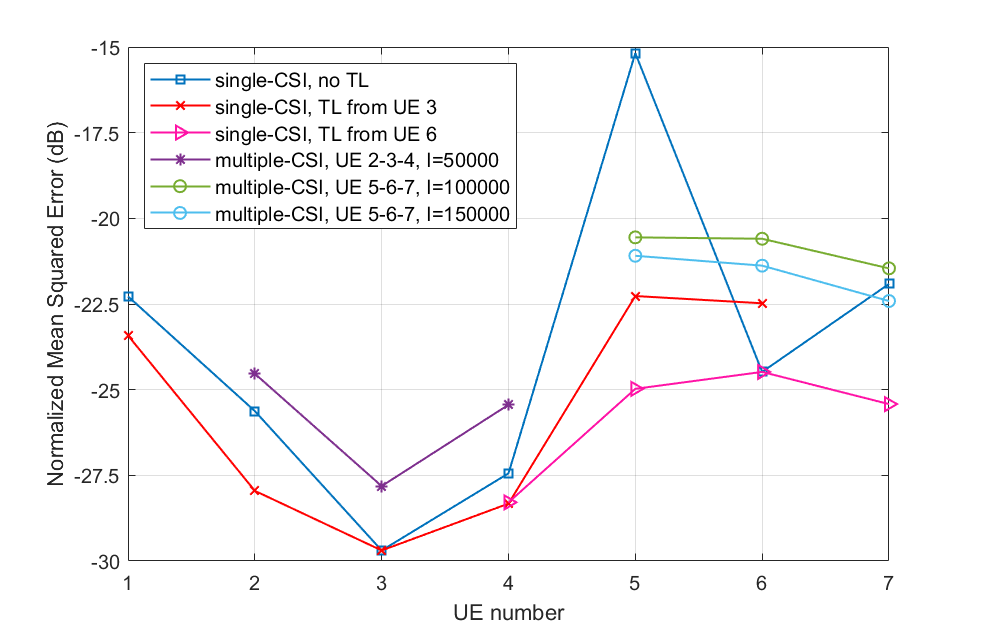}
    \caption{Results for the single-CSI estimation with transfer learning and simultaneous CSI recreation for multiple UEs with channel measurements at SNR~$=20$~dB.}
    \label{fig:transfer}
\end{figure}

\begin{figure}[!tb]
    \centering
    \includegraphics[width=\columnwidth]{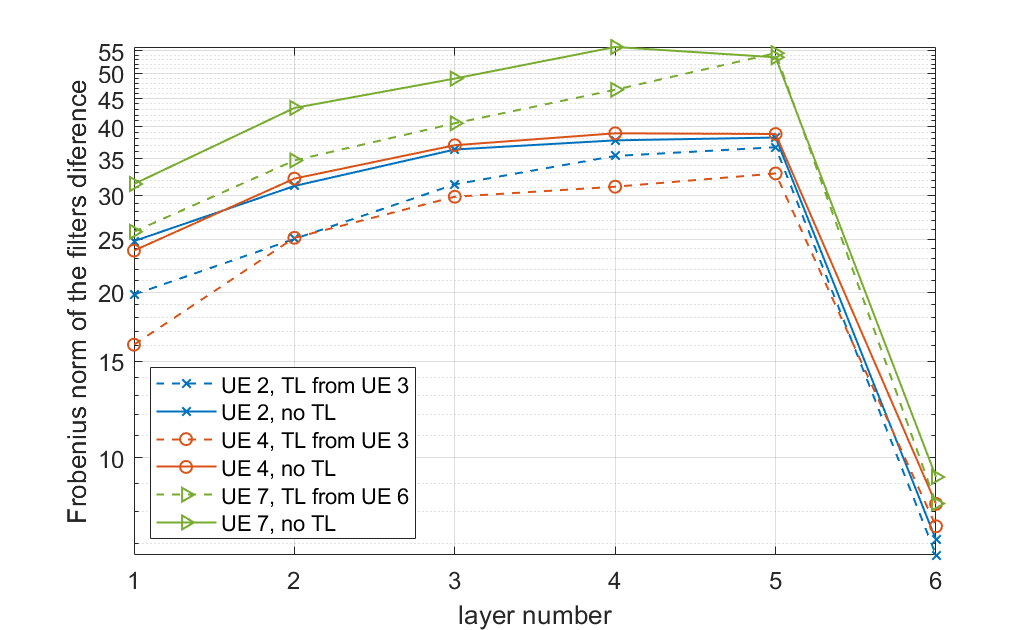}
    \caption{Frobenius norm difference of the convolutional filters between each layer. Filters derived from random initialization are in complete line, while filters from transfer learning are in dashed lines.}
    \label{fig:normF}
\end{figure}

Figure~\ref{fig:transfer} presents the results of transfer learning (TL) for single UE CSI estimation at measurement SNR~$=20$~dB. 
For reference, we plot in blue the results for each UE with random initialization. 
As UE~$3$ has the best estimation gain, we take its weights as starting point for optimizing the weights $\mathbfcal{K}^*_{2,3}$ and $\mathbfcal{K}^*_{4,3}$ for UE~$2$ and UE~$4$, respectively. 
After that, we estimate UE~$1$ initializing from $\mathbfcal{K}^*_{2,3}$ and UE~$5$ initializing from $\mathbfcal{K}^*_{4,3}$, and so on. This propagation of transfer learning from UE~$3$ is plotted in red. 
We can see that the transfer learning approach was successful to find a solution $\mathbfcal{K}_{5,4-3}^*$ that provides CSI reconstruction within our design requirements. 
As the channel gain for UE~$6$ estimated with transfer learning from UE~$3$ is smaller than the estimation gain of UE~$6$ starting from random initialization, we take UE~$6$ as a second basis for transfer learning. 
The transfer learning results for UEs $5$ and $7$, and then UE~$4$ are plotted in pink. Hence, UE~$6$ is a better transfer learning basis for UEs $5$ and $7$. 
However, it does not provide advantage for UE~$4$. 
In Figure~\ref{fig:normF} we plot the Frobenius norm of the difference between the filters $\mathbfcal{W}_l$ in each layer when initialized from random values (no TL) and when using initialization from the neighbor's weights. 
Those results indicates that equation~\ref{eq:filternorm} is correct, the derived sub-spaces ($\mathbfcal{K}^*_{2,3}$, $\mathbfcal{K}^*_{4,3}$, and $\mathbfcal{K}^*_{7,6}$) where constrained to be close to the initialization sub-spaces ($\mathbfcal{K}^*_{3,0}$ and $\mathbfcal{K}^*_{6,0}$). 
For reporting the optimal weights, the worst case requires transmission of all parameters per UE, which is only $17.45\%$ of the full channel coefficients. 
However, if the filters are closer to each other, differential compression schemes can be applied to further reduce the number of parameters to be reported.

Based on the previous results, we set two candidate UE-groups for simultaneous CSI recreation. 
We derive a UNN architecture for estimating simultaneously UEs $2,3$ and $4$, and a second architecture for UEs $5, 6$ and $7$.
For UEs $2,3$ and $4$, the UNN architecture has $L=6$ layers, 3D convolutional filters with $k_{1:L-2} = 64$, and upsampling operation only in the subcarriers and time snapshots dimensions ($\mathbf{D}_l$ is deactivated). 
Hence, $\mathbfcal{Z}_0^M~\in~\mathbb{R}^{4 \times 4 \times 3 \times 64}$ is drawn from a uniform distribution $U(-0.15,+0.15)$ and the number of gradient iterations is $I=50000$. The estimation error is presented in purple at Figure~\ref{fig:transfer}. 
This architecture meets the design goal, but provides less estimation gain than the UNN estimator with transfer learning.
However, since the convolutional filters are of size one, the number of trainable parameters for multiple CSI recreation of UEs $2, 3$ and $4$ is the same as for single UE CSI recreation. 
This means that, we achieve a higher compression rate as the number of trainable parameters corresponds to just $5.82\%$ of the coefficients in $\mathbfcal{H}_\mathrm{mes}^{C~3}~\in~\mathbb{C}^{64 \times 64 \times 3 \times 36}$.

For multiple CSI recreation of UEs $5,6$ and $7$, there are $L=7$ layers from which $4$ are inner layers with upsampling $\mathbf{D}_l$ deactivated, and $2$ pre-output layers with hyper-parameters $k_{1:L-1}=64$. 
The random seed is $\mathbfcal{Z}_0^M~\in~\mathbb{R}^{4 \times 4 \times 3 \times 64}$ and $I=100000$ iterations for gradient update. 
We plot the estimation result in Figure~\ref{fig:transfer} using a green line. 
There is about $4$~dB difference between the multiple-CSI estimator and the transfer learning estimator. 
Compared with the multiple-CSI recreation for UEs $2, 3$ and $4$, we need more parameters ($k$) and more iterations. 
Nonetheless, the architecture for multiple-CSI recreation of UEs $5, 6$ and $7$ needs only $6.77\%$ of the number of coefficients in $\mathbfcal{H}_\mathrm{mes}^{C~3}$. 
We change the number of iterations $I=150000$ for the multiple-CSI recreation of UEs $5, 6$ and $7$. 
The result is plotted in light blue on Figure~\ref{fig:transfer}. 
There is a further $1$~dB gain, but the convergence is very slow (50k iterations to improve just $1$~dB).      
This indicates that the simultaneous channel estimation for UEs $5, 6$ and $7$ is more challenging.

\section{Conclusion}
\label{sec:conclusion}

In this paper we propose to use transfer learning to take advantage of the prior knowledge stored in the UNN structure. 
Moreover, we present a UNN architecture for simultaneous CSI estimation for multiple UEs which can further reduce the number of trainable parameters. 
In addition, we show the compression benefits of UNN structures which can further leverage low-overhead CSI reporting. 
Our results show that the UNN structure is able to inherently learn the environment characteristics when fitting the measured channels. 
By transfer learning, we are able to access this prior knowledge and have a higher channel estimation gain. 
Due to the channel correlation between neighboring UEs, we can
simultaneously estimate CSI for multiple UEs with a $3$-d kernel UNN architecture that reduces the number of trainable parameters at the price of smaller channel estimation gain.
Future work may consider the use of compression schemes on top of UNN structures to increase the compression rate.

\section*{Acknowledgement}
This research was partly funded by German Ministry of Education and Research (BMBF) under grant 16KIS1184 (FunKI).

\bibliographystyle{IEEEtran}
\bibliography{bibfile}

\end{document}